\documentclass[aps,prc,nofootinbib,letterpaper,showkeys,%F
superscriptaddress,amsmath,amssymb,showpacs,floatfix,%
twocolumn]{revtex4}
\usepackage{bm}
\usepackage{graphics,graphicx}
\usepackage{amssymb}
\begin{document}
\newcommand{\be}{\begin{equation}}
\newcommand{\ee}{\end{equation}}
\newcommand{\anu}{\ensuremath{\bar{\nu}}}
\newcommand{\anue}{\ensuremath{\bar{\nu}_e}}
\newcommand{\rE}{\ensuremath{R_{\oplus}}}
\newcommand{\nucleus}[2][]{\ensuremath{{}^{#1}\mathrm{#2}}}
\newcommand{\U}[1][]{\nucleus[#1]{U}}
\newcommand{\Th}[1][]{\nucleus[#1]{Th}}
\newcommand{\K}[1][]{\nucleus[#1]{K}}
\newcommand{\Rb}{\nucleus[87]{Rb}}
\newcommand{\eg}{\emph{e.g.},\ }
\newcommand{\ie}{\emph{i.e.\ }}
\newcommand{\et}{\emph{et al.\ }}
\title{Nuclear physics for geo-neutrino studies}
\thanks{Corresponding author: Marcello Lissia}

\author{Gianni Fiorentini}
%\email{fiorenti@fe.infn.it}
\affiliation{Dipartimento di Fisica,
Universit\`a degli Studi di Ferrara, I-44100 Ferrara, Italy}
\affiliation{Istituto Nazionale di Fisica Nucleare, Sezione di
Ferrara, I-44100 Ferrara, Italy}
\author{Aldo Ianni}
%\email{aldo.ianni@lngs.infn.it}
\affiliation{Istituto Nazionale
di Fisica Nucleare, Laboratori Nazionali del Gran Sasso,
             I-67010 Assergi(AQ), Italy}

\author{George Korga}
%\email{george.korga@lngs.infn.it}
\affiliation{Istituto Nazionale
di Fisica Nucleare, Laboratori Nazionali del Gran Sasso,
             I-67010 Assergi(AQ), Italy}

\author{Marcello Lissia}
\email{marcello.lissia@ca.infn.it}
\affiliation{Istituto Nazionale
di Fisica Nucleare, Sezione di Cagliari,
             I-09042 Monserrato, Italy}
%\affiliation{Dipartimento di Fisica, Universit\`a di Cagliari,
%             I-09042 Monserrato, Italy}
%
\author{Fabio Mantovani}
%\email{mantovani@fe.infn.it}
% \email{fabio.mantovani@unisi.it}
\affiliation{Dipartimento di Fisica,
Universit\`a degli Studi di Ferrara, I-44100 Ferrara, Italy}
\affiliation{Istituto Nazionale di Fisica Nucleare, Sezione di
Ferrara, I-44100 Ferrara, Italy}
\affiliation{Centro di GeoTecnologie CGT,I-52027 San Giovanni
Valdarno, Italy}

\author{Lino Miramonti}
%\email{}
 \affiliation{Dipartimento di Fisica, Universit\`a degli Studi di
Milano, I-20133 Milano,  Italy}
\affiliation{Istituto Nazionale di Fisica Nucleare, Sezione di
Milano, I-20133 Milano, Italy}

\author{Lothar Oberauer}
%\email{}
 \affiliation{Physik Department, Technische Universit\"at Muenchen, 85747 Garching, Germany}

\author{Michel Obolensky}
%\email{}
 \affiliation{Laboratoire AstroParticule et Cosmologie, 75231 Paris cedex 13, France}

\author{Oleg Smirnov}
%\email{oleg.smirnov@lngs.infn.it}
 \affiliation{Joint Institute for Nuclear Research, 141980, Dubna, Russia}

\author{Yury Suvorov}
%\email{}
\affiliation{Istituto Nazionale
di Fisica Nucleare, Laboratori Nazionali del Gran Sasso,
             I-67010 Assergi(AQ), Italy}
%

%\date{\today}
\date{August 24th, 2009}
\begin{abstract}
Geo-neutrino studies are based on theoretical estimates of geo-neutrino spectra.
We propose a method for a direct measurement of the energy distribution of antineutrinos
from decays of long-lived radioactive isotopes. We present preliminary results for the
geo-neutrinos from \nucleus[214]{Bi} decay, a process which accounts for about one half of the total
geo-neutrino signal. The feeding probability of the lowest state of \nucleus[214]{Bi} --- the most
important for geo-neutrino signal --- is found to be
$p_0= 0.177 \pm 0.004 \text{\ (stat) } {}^{+0.003}_{-0.001} \text{\ (sys)}$,
under the hypothesis of Universal Neutrino Spectrum Shape (UNSS).
This value is consistent with the (indirect) estimate of the Table of Isotopes (ToI).
We show that achievable larger statistics and reduction of systematics should
allow  to test possible distortions
of the neutrino spectrum from that predicted using the UNSS hypothesis.
Implications on the geo-neutrino signal are discussed.
\end{abstract}

%\pacs{{\bf 91.35.-x, 13.15.+g, 14.60.Pq, 23.40.Bw}}
\maketitle

%%%%%%%%%%%%%%%%%%%%%%%%%%%%%%%%%%%%%%%%%%%%%%%%%%%%%%%%%%%%%%%%%%%%%%
%\tableofcontents

\section{\label{sec:intro}Introduction}
Geo-neutrinos, the antineutrinos from the progenies of \U, \Th\ and \K[40]
decays in the Earth, bring to the surface information from the whole planet,
concerning its content of radioactive elements. Their detection can shed light
on the sources of the terrestrial heat flow, on the present composition,
and on the origins of the Earth.

Although geo-neutrinos were conceived very long ago, only recently they have been
considered seriously as a new probe of our planet interior, as a consequence of two
fundamental advances that occurred in the last few years: the development of large
extremely-low-background neutrino detectors and the progress on understanding neutrino
propagation. From the theoretical point of view, the links between the geo-neutrino
signal and the total amount of natural radioactivity in the Earth have been analyzed
by several groups. Various ``reference models''
 \cite{Mantovani:2003yd,Fogli:2006zm,Enomoto:2007}
for geo-neutrino production have
been presented in the literature; some of these models have been refined with geological
and geochemical studies of the regions surrounding the detectors \cite{Fiorentini:2005cu}.
KamLAND \cite{Araki:2005qa,Abe:2008ee} and
Borexino \cite{Alimonti:2000xc,Alimonti:2008gc}
are collecting geo-neutrino data, while several planned experiments
(\eg SNO+, LENA, HANOHANO, EARTH, \ldots) have geo-neutrino measurements among
their primary goals. A recent review is presented in \cite{Fiorentini:2007te}.

This activity has to be complemented with some deepening of the nuclear physics that
is at the basis of geo-neutrino detection and which is crucial for interpreting future
geo-neutrino data. The aim of this paper is to discuss the uncertainties of some nuclear
physics parameters that enter in the interpretation of the geo-neutrino signal and to
provide a framework for an experimental determination of these parameters.

In all experiments that use hydrocarbons as detection media,
either running or in preparation, the reaction for geo-neutrino
detection is the inverse beta decay on free protons
\be
\label{eq:inverseBeta}
\anue + p \to e^+ + n - 1.806\mathrm{\ MeV} \quad .
\ee
The signal is estimated from the cross section  $\sigma(E_{\nu})$ of
Eq.~(\ref{eq:inverseBeta}) and from the decay spectra $f(E_{\nu})$ of geo-neutrinos produced
in each beta decay along the decay chains, the relevant quantity being the
``specific signals'' defined as:
\be
\label{eq:specificSignal}
s_i = \int_{E_0}^{E_{\mathrm{max}}} dE_{\nu} \sigma(E_{\nu}) f_i(E_{\nu})\quad ,
\ee
where $E_0 = 1.806$~MeV is the threshold energy for reaction~(\ref{eq:inverseBeta}),
$E_{\mathrm{max}}$ is the maximal geo-neutrino energy and the spectrum
$f(E_{\nu})$ is normalized to one  geo-neutrino~\footnote{A detector with
$N_p$ free protons will collect a signal rate $S = N_p \sum_i \Phi_i s_i $,
where $\Phi_i$ are the incoming fluxes of geo-neutrinos from the $i$-th beta
decay in the chain and $s_i$ are the corresponding specific signals.}
\be
\label{eq:specificSignalNorm}
\int_{0}^{E_{\mathrm{max}}} dE_{\nu} f(E_{\nu}) =1 \quad .
\ee

It is important to observe that the specific signal is affected by unknown uncertainties.
In fact, whereas  $\sigma(E_{\nu})$ is affected by uncertainties of less than percent
\cite{Vogel:1999zy,Strumia:2003zx},  it is difficult to assess the accuracy of
$f(E_{\nu})$, which is determined from rather indirect measurements and questionable
 theoretical assumptions.

\emph{Our goal is to provide a framework for a direct measurement of  $f(E_{\nu})$,
 so that the accuracy of the specific signal can be established.}

\section{\label{sec:Why}Why should geo-neutrino spectra be measured?}
Geo-neutrinos are produced through pure $\beta$  and $\beta$-$\gamma$ processes:
\begin{eqnarray*}
X &\to& X' + e + \anue  \\
X &\to& X'^*+ e + \anue   \\
  &&  \quad \searrow  \\
  && \quad\quad X' + n + \gamma \quad .
\end{eqnarray*}
In order to determine the geo-neutrino decay spectra $f(E_{\nu})$ one has to know:
\begin{enumerate}
  \item[(i)] the feeding probabilities $p_n$ of the different energy states of the final nucleus;
  \item[(ii)] the shape of the neutrino spectrum for each transition.
\end{enumerate}
Let us discuss in some detail the procedures and assumptions used for deriving these quantities.

Feeding probabilities are derived from measurements of the intensities $I^{m,n}_{\gamma}$
of the gamma lines. These are corrected for internal conversion in order to derive the transition
probabilities from level $m$ to $n$:
\be
I^{m,n} =I^{m,n}_{\gamma} (1+\alpha^{m,n}) \quad .
\ee
The internal conversion coefficients  $\alpha^{m,n}$ are obtained by theoretical calculations.
In general they are of order $10^{-2}$, unless selection rules forbid or inhibit the gamma
emission~\footnote{An important case in this respect is the
$E0$ transition of \nucleus[214]{Bi} at 1415.8~keV, which occurs essentially through internal
conversion.}.

The feeding probabilities for the excited states are then obtained with a subtraction procedure
as the difference between the intensities of outgoing and ingoing transitions:
\be
p_n=  \sum_{m<n} I^{n,m} -  \sum_{m>n} I^{m,n}   \quad .
\ee
The feeding probability of the lowest state, $p_0$, is obtained with the same subtraction procedure:
\be
        p_0 = 1  -  \sum_{m>0} I^{m,0}  \quad .
\ee
This procedure implies that all transitions to the ground state that are not observed,
or taken into account, are included in the feeding probability to the lowest energy state.
\emph{In other words, $p_0$ is indirectly determined,
 whereas it is of special interest for our purposes}:
$\beta$ transitions directly to the lowest energy state,
the ``pure $\beta$'', produce the most energetic geo-neutrinos and thus give the largest
contribution to the specific signal.

For each transition, the shape of the neutrino spectrum is generally calculated
assuming the well-known ``universal shape'' distribution. This expression,
see~\cite{Fiorentini:2007te}, corresponds to momentum independent nuclear
matrix elements (as for allowed transitions) and includes the effect of
the bare Coulomb field of the nucleus through the relativistic Fermi function.
Electron screening and nuclear finite size effects are not considered.
Note that this same "universal shape" expression is used even for the forbidden
transitions
(see Tables~\ref{tab:uBetaTransitions} and \ref{tab:thBetaTransitions}),
where momentum-dependent nuclear matrix elements can appear.

\begingroup
\squeezetable

\begin{table*}[htb] \caption[eee]{Effective transitions in the \U[238] chain
from~\cite{Fiorentini:2007te}. In bold the reactions that give most of the signal.
For each decay the table shows the
probability, the maximal antineutrino energy,
the intensity $I_k$, its error $\Delta I_k$, type and percentage contributions to the uranium
geo-neutrino signal,  and to the  (\U + \Th) geo-neutrino signal.
For this last column it is assumed the chondritic ratio for the masses
($\Th/\U= 3.9$), which implies that 79\% of the geo-neutrino signal
comes from uranium. \label{tab:uBetaTransitions}
}
\newcommand{\dg}{\hphantom{$0$}}
\renewcommand{\tabcolsep}{1pc} % enlarge column spacing  prima era 2pc
\begin{tabular}{cccccccc}
 \hline\hline
    $i\to j $  & $R_{i,j}$ & $E_{max}$  & $I_k$ & $\Delta I_k$ & Type & $S_{\U}$ & $S_{\mathrm{tot}}$ \\
            &    &      [keV] & & & [\%] & [\%]\\
 \hline
 $ \nucleus[234]{Pa}_m\to  \nucleus[234]{U} $ & \textbf{0.9984} & \textbf{2268.92} & \textbf{0.9836} & \textbf{0.002\dg }& \textbf{\dg 1st forbidden} $(0^-) \to 0^+$  & \textbf{39.62} & \textbf{31.21} \\
\hline
   $ \nucleus[214]{Bi}\to  \nucleus[214]{Po}  $ & \textbf{0.9998 }& \textbf{3272.00 }   & \textbf{0.182\dg} & \textbf{0.006\dg }&\textbf{ 1st forbidden} $1^- \to 0^+$ &  \textbf{ 58.21} &   \textbf{45.84} \\
                                                &        & 2662.68    & 0.017\dg & 0.006\dg & 1st forbidden $1^- \to 2^+$ & \dg1.98 & \dg1.55 \\
                                                &        & 1894.32    & 0.0743   & 0.0011   & 1st forbidden $1^- \to 2^+$ & \dg0.18 & \dg0.14 \\
                                                &        & 1856.51    & 0.0081   & 0.0007   & 1st forbidden $1^- \to 0^+$ & \dg0.01 & \dg0.01 \\
   \hline \hline
\end{tabular}\\[2pt]
\end{table*}
\endgroup
\begingroup
\squeezetable
\begin{table*}[htb] \caption[jjj]{Effective transitions in the \Th[232] chain
from~\cite{Fiorentini:2007te}.
In bold the reaction that gives most of the signal.
For each decay the table shows the
probability, the maximal antineutrino energy, the
intensity $I_k$, its error $\Delta I_k$, type and percentage contributions to the thorium
geo-neutrino signal, and to the total (\U + \Th) geo-neutrino
signal. For this last column it is assumed the chondritic ratio for the
masses ($\Th/\U= 3.9$), which implies that 21\% of the geo-neutrino
signal comes from thorium. \label{tab:thBetaTransitions} }
\newcommand{\dg}{\hphantom{$0$}}
\renewcommand{\tabcolsep}{1pc} % enlarge column spacing
\begin{tabular}{cccccccc}
 \hline\hline
    $i\to j $  & $R_{i,j}$ & $E_{max}$  & $I_k$ & $\Delta I_k$ & Type & $S_{\Th}$ & $S_{\mathrm{tot}}$ \\
            &    &      [keV] & & & [\%] & [\%]\\
 \hline
 $ \nucleus[212]{Bi}\to  \nucleus[212]{Po} $    & \textbf{0.6406} & \textbf{2254\dg\dg} & \textbf{0.8658} & \textbf{0.0016 }& \textbf{1st forbidden } $1^{(-)} \to 0^+$  & \textbf{94.15} & \textbf{20.00} \\
\hline
   $ \nucleus[228]{Ac}\to  \nucleus[228]{Th}  $ & 1.0000 & 2069.24    & 0.08\dg\dg & 0.06\dg\dg& Allowed $3^+ \to 2^+$ & \dg5.66 & \dg1.21 \\
                                                &        & 1940.18    & 0.008\dg & 0.006\dg & Allowed $3^+ \to 4^+$ & \dg0.19 & \dg0.04 \\
   \hline \hline
\end{tabular}\\[2pt]
\end{table*}

\endgroup

These observations suggest that the feeding probabilities need to be confirmed by
different experimental techniques and the electron decay spectrum
need to be experimentally tested.

\section{\label{sec:decaySpectra}Towards a direct measurement of geo-neutrino decay spectra}
\begin{figure}
\includegraphics[width=0.6\columnwidth,angle=-90]{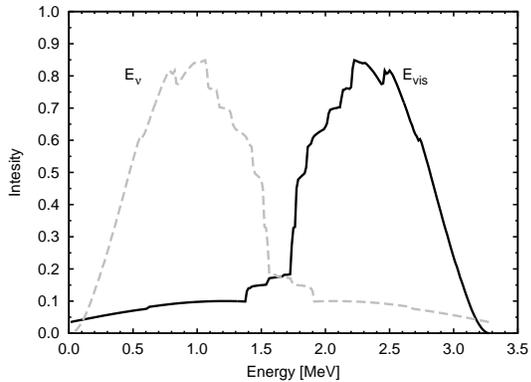}
 \caption[aaa]{The decay spectrum of \nucleus[214]{Bi} as a function of the visible
  energy $E_{\mathrm{vis}}$ (full line) and of the neutrino energy $E_{\nu}$  (dashed line).
\label{fig:decaySpectrum}
}
\end{figure}

When the nucleus $X$ decays, whichever is the transition involved, energy conservation
provides a connection between the neutrino energy $E_{\nu}$, the kinetic energy of the
electron $T_e$, and the total energy of the emitted gammas, $E_{\gamma}$:
\be
Q = E_{\nu} + T_e + E_{\gamma}\quad ,
\ee
where $Q = M_{X} -M_{X'} -M_{e}$ is the $Q$-value for the decay.
In order to measure the geo-neutrino spectrum, one needs a
calorimetric detector which is capable
of measuring the ``visible'' energy deposited together by
electrons~\footnote{Note that the energy deposited by conversion electrons is also included.}
and gammas, $ E_{\mathrm{vis}} = T_e + E_{\gamma}$.
 \emph{When measured decay events are displayed as a function of $ E_{\mathrm{vis}}$,
  by a mirror reflection one immediately obtains the number of events as a function
  of neutrino energy, at  $E_{\nu} = Q - E_{\mathrm{vis}}$};
  as an example, see figure~\ref{fig:decaySpectrum}
  for the decay spectrum of \nucleus[214]{Bi}.

  %% figura 1

  For such a measurement one needs a detector that can collect the energy lost by both
  electrons and gammas and that has a similar response to both particles.
  Essentially, this is a calorimetric measurement. In principle, it can be done
  with large bolometers~\cite{gorlaPrivate},
  which have very good energy resolution but long dead times.
  A sufficiently large liquid scintillator detector is suitable for such measurements.
  Although energy resolution is limited, nevertheless it can contain both electrons and gammas
  and a significant statistics can be collected in a reasonable time.

\begin{figure}
\includegraphics[width=0.6\columnwidth,angle=-90]{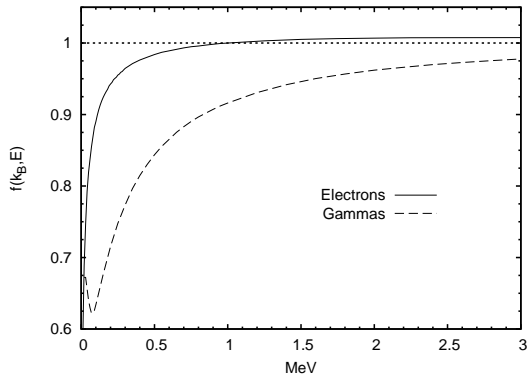}
 \caption[bbb]{Quenching factor for electrons and gammas, for
 a Birks coefficient $k_B = 1.95\cdot 10^{-3}$cm~MeV$^{-1}$, see appendix.
 The quenching factor has been normalized so that is one for 1~MeV electrons.
\label{fig:quenchingFactor}
}
\end{figure}

There are some limitations that should be considered,
 when using a scintillator as a calorimeter.
An ideal detector should provide the same response for gammas and electrons with equal energy,
independently of the positions where the particles are generated. In practice, however:
\begin{enumerate}
  \item[(i)] even in a very large detector, the energy released as scintillation light from electrons
  and gammas of the same energy are not the same.
  This difference becomes marked at low energy,
  see Fig.~\ref{fig:quenchingFactor}.
  \item[(ii)] Gammas can escape from a finite detector, thus releasing only a fraction of their energy.
  \item[(iii)] The number of photons collected by the detector can depend on the position where
  they have been produced (due to absorption, optical coverage, \ldots).
\end{enumerate}

%% figura 2

All these effects can be taken into account by using calibration measurements,
by selecting events that occur in the inner part of the detector (in order to minimize corrections
due to escaping gammas) and with energy above a suitable threshold. The comparison between experimental
spectra and theoretical predictions has to be implemented by means of a
Montecarlo simulation that accounts for the actual characteristics of the detector.

\section{\label{sec:detector}The proposed detector}
We propose to exploit the potential of the Counting Test Facility (CTF),
which is operational and available in the underground I.N.F.N. Gran Sasso
National Laboratory.

\begin{figure}
\includegraphics[width=0.6\columnwidth,angle=0]{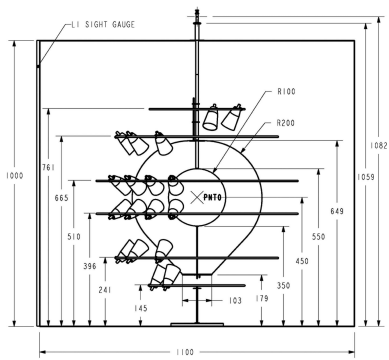}
 \caption[ccc]{Side view of the design of CTF. The vessel
 (labelled R100 in this drawing) and shroud (R200) are shown,
 as well as the six rings of PMT's, the cylindrical tank, and the
 tubes used for filling and draining the vessel.
 The point PNT0 is the nominal center of the sphere of PMT's and of
 the CTF vessel. Dimensions are given in cm. Courtesy of the Borexino collaboration.
\label{fig:CTFdesign}
}
\end{figure}

\begin{figure}
\includegraphics[width=0.6\columnwidth,angle=0]{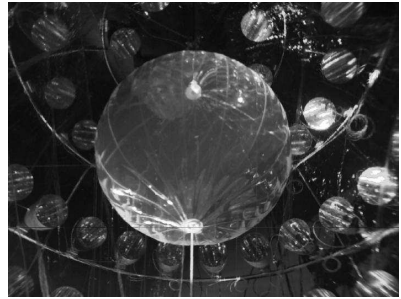}
 \caption[ddd]{A picture of the CTF viewed from below.
 Courtesy of the Borexino collaboration.
\label{fig:CTFphoto}
}
\end{figure}

Like Borexino, the CTF design~\cite{Alimonti:1998nt} is based on the principle
of graded shielding, see figures \ref{fig:CTFdesign} and \ref{fig:CTFphoto}.
The active scintillation liquid in CTF is
a four-ton mass of pseudocumene enclosed in a transparent nylon sphere, the CTF vessel.
Outside this vessel there is a volume of ultra-pure water which is enclosed in a second
nylon sphere, the so-called CTF radon shroud, intended to prevent radon transport
with thermal fluxes from the outside zones of the detector.
A set of inward-facing PMTs is arrayed outside the shroud.
The entire apparatus, surrounded by another volume of water, is contained in a cylindrical stainless tank.
The bottom surface of the tank holds 16 upward-facing PMTs used to tag the
muons passing through the detector by means of the their Cherenkov light
in the water.

%% figura 3

%% figura 4

 The facility is equipped with a rod system, which can be used to insert a small,
 cylindrical quartz vial inside the CTF vessel. A suitable source, dissolved in the
 liquid scintillator, can be placed in the vial. Electrons are stopped inside the
 vial and the scintillation light is propagated within CTF through the quartz
 (which is transparent to the near-UV wavelengths of scintillation light and has
 index of refraction close to that of the scintillator) whereas gamma conversion
 occurs inside the CTF inner vessel. The inward facing PMTs outside the shroud can
 thus detect light originating from both electrons and gammas.

A Montecarlo code has been developed for CTF. It is described in~\cite{Back:2004hd}
and in the Appendix with the adjustments for our specific task.

\section{\label{sec:whatMeasure}What has to be measured?}
Geo-neutrinos with energy above the threshold for reaction~(\ref{eq:inverseBeta}) arise
only from the chains of \U[238] and \Th[232].

In particular, for \U[238] only three nuclides (\nucleus[234]{Pa}, \nucleus[214]{Bi},
\nucleus[210]{Tl})
contribute to the geo-neutrino signal.
The contribution from \nucleus[210]{Tl} is negligible, due to its small occurrence
 probability, and the uranium contribution to the geo-neutrino signal comes from
 five $\beta$ decays: one from \nucleus[234]{Pa} and four from \nucleus[214]{Bi}.
  Table~\ref{tab:uBetaTransitions}
  lists the effective transitions, \ie, those that can produce
 antineutrinos with energy above the threshold $E_0$.
 In fact, 98\% of the uranium signal arises from the two transitions to the ground state
  (in bold in Table~\ref{tab:thBetaTransitions})
   and an accuracy better than 1\% is achieved by adding the third one.

\nucleus[232]{Th} decays into \nucleus[208]{Pb} through a chain of
six $\alpha$-decays and four $\beta$-decays. In secular equilibrium the complete network
includes five $\beta$-decaying nuclei. Only two nuclides (\nucleus[228]{Ac}
 and \nucleus[212]{Bi}) yield antineutrinos with energy larger than 1.806~MeV.
 The thorium contribution to the geo-neutrino signal comes from three $\beta$ decays:
 one from \nucleus[212]{Bi} and two from \nucleus[228]{Ac}
 (see Table~\ref{tab:thBetaTransitions}).
 In fact, 94\% of the thorium signal arises from the transition to the ground state
 of  \nucleus[212]{Po} (in bold in Table~\ref{tab:thBetaTransitions}).

We remind that, assuming the chondritic ratio for the global uranium and thorium
mass abundances, $a(\Th)/a(\U)=3.9$, one expects that geo-neutrinos
from uranium (thorium) contribute about 80\% (20\%) of the total $\U + \Th$ geo-neutrino signal.

In summary:
\begin{enumerate}
  \item[(a)]  98\% of uranium geo-neutrino signal comes from just two transitions,
  one from \nucleus[214]{Bi} and the other from \nucleus[234]{Pa}.
  They provide 77\% of the expected total $\U + \Th$ signal.
  \item[(b)]  A single decay of \nucleus[212]{Bi} accounts for 94\% of the
  thorium signal. It provides 20\% of the expected $\U + \Th$ signal.
\end{enumerate}
Just three transitions have to be investigated experimentally.
In this respect, the following considerations can be useful:
\begin{enumerate}
  \item[(a)] \nucleus[222]{Rn} ($\tau_{1/2} = 3.824$~days) can be easily dissolved in
  the scintillator and the decay of \nucleus[214]{Bi} is uniquely identified by the
  subsequent decay of \nucleus[214]{Po} ($\tau_{1/2} = 164.3 \mu$s).
  \item[(b)] By dissolving \U[238] in the scintillator, one can detect the beta decay
  of \nucleus[234]{Pa} (superimposed, however, with that of \nucleus[234]{Th}).
  The subsequent decays of the chain are effectively blocked by the long half-live
  of \U[234] ($\tau_{1/2} = 2.455\cdot 10^{5}$~yr).
  \item[(c)] For the investigation of \nucleus[212]{Bi} decay one has to start with
  a \nucleus[224]{Ra} source ($\tau_{1/2} = 3.66$~days) or with a \nucleus[232]{Th} source.
  The decay of \nucleus[212]{Bi} can be easily identified by the subsequent $\alpha$ decay of
  \nucleus[212]{Po} ($\tau_{1/2} = 299$~ns).
\end{enumerate}

\section{\label{sec:results}Results from a diffuse Rn source}
In order to test the method that we are proposing, we have used data from a sizeable,
though limited, radon contamination of CTF, which occurred in the early phase of
operation of the
detector. In the full volume of CTF, we selected the candidates
$\beta$-decay from \nucleus[214]{Bi} by the distinctive subsequent \nucleus[214]{Po}
$\alpha$ decay, which occurs with a mean time-delay of $237\, \mu$s (the so-called BiPo events).

\begin{figure}
\includegraphics[width=0.6\columnwidth,angle=-90]{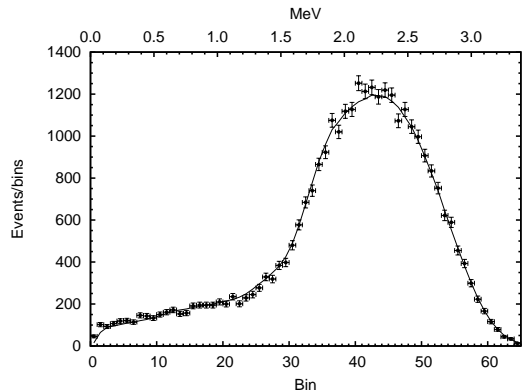}
 \caption[eee]{Data and best fit. Events have been grouped
 into 65 bins from 0 to 3.4~MeV. Event numbers together with statistical
 error (vertical bar) and bin size (horizontal bar) are presented as a
 function of bin number, see text. We only fit data in bins from 3 to 65.
 The continuous curve shows the best fit when three parameters
 ($p_0$, light yield, and normalization) are left as free.
\label{fig:dataBestFit}
}
\end{figure}

The selection of the analyzed events, see the data
points in Fig.~\ref{fig:dataBestFit}, is described in the following two
subsections
\ref{subsec:dataSelection} and \ref{subsec:dataAnalysis}.

 %% figura 5

\begin{figure}
\includegraphics[width=0.6\columnwidth,angle=0]{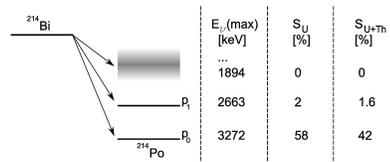}
 \caption[fff]{A simplified decay scheme of \nucleus[214]{Bi}
  and the contribution of the various levels to the total geo-neutrino signal.
\label{fig:biDecayScheme}
}
\end{figure}

We recall that most of the contribution to the geo-neutrino signal arises from the transition
to the lowest energy state (0) of \nucleus[214]{Po}, see Fig.~\ref{fig:biDecayScheme}.
 Our analysis aims:
\begin{enumerate}
  \item[(1)] to determine the probability $p_0$ of populating the lowest energy  state
  (assuming the universal ``allowed'' shape) from the CTF data.
  \item[(2)] To determine whether the spectrum of the pure beta transition (that to the lowest state)
  is deformed with respect to the universal allowed shape.
  \item[(3)] To discuss the implications of this study on the specific
  geo-neutrino signal $s(\nucleus[214]{Bi})$, given by
  Eq.~(\ref{eq:specificSignal}).

\end{enumerate}
%% figura 6

\subsection{\label{subsec:dataSelection}Data selection and backgrounds}
The main selection criterion is that the coincidence time between consecutive signals,
provided by the prompt  $\beta$ decay of \nucleus[214]{Bi} and the delayed  $\alpha$
decay of \nucleus[214]{Po}, be $2  \mu s < \Delta t_{1/2} < 602 \mu s$.
The lower limit eliminates fast coincidences from the corresponding decays in
the \nucleus[232]{Th} chain, whereas the higher limit has been chosen to
keep random coincidences under 1\%, while preserving high statistics.
The selected events are $4.54 \cdot 10^5$.

Then we require that the energy deposited by the first signal is $E_1 < 3.9$~MeV,
taking into account the $Q$-value ($Q=3.27$~MeV) and the energy resolution,
about 0.2~MeV at these energies.
This cut removes random coincidences, while keeping the Bi-Po events
(acceptance almost 100\%).

The reconstructed radial
positions~\footnote{A typical spatial $1\sigma$  resolution is of order 10 cm at 1 MeV.}
of the two signals $r_1$ and $r_2$, are used to impose the three
conditions: $r_1 < 2$~m, $r_2 < 2$~m, and $|r_1 - r_2|< 2.5$~m.
These very weak cuts have total acceptance efficiency about 100\%,
while removing random coincidences.

We then impose that the energy of the second signal, on the electron energy scale,
is $0.56 \mathrm{\ MeV} < E_2 < 1.1$~MeV (note that the 7.9~MeV alpha particles
from \nucleus[214]{Po} decay are quenched by a factor about 11), in order to
reduce low-energy  $\alpha$'s of the Rn chain and random coincidences.
The acceptance efficiency is 98.7\%.

CTF has good pulse-shape discrimination between $\alpha$ and $\beta$
events~\cite{Back:2007gp}.
To avoid contamination by high-energy  $\alpha$'s, we add cuts on a
suitable $\alpha / \beta$ discrimination parameter.
The combined acceptance efficiency is 99.4\% and we are left
with $4.46 \cdot 10^5$ events.

The remaining background is estimated by applying the same sets of cuts,
with the coincidence time window $2000  \mu\mathrm{s} < \Delta t_{1/2} <8000 \mu$s.
In this way, we estimate that contamination of random coincidences is about 0.7\%.

\subsection{\label{subsec:dataAnalysis}Data analyses and fiducial region}
Taking into account the number of events and the estimated energy resolution
$\Delta E \approx 80 \mathrm{\ keV} \cdot (E / \mathrm{MeV})^{1/2}$,
we grouped events in  bins of about 50-keV~\footnote{The energy scale depends
on the best-fit value of the light yield. At the best fist the bin size is 52~keV.}
and analyzed the 63 bins from 0.1 MeV up to 3.4~MeV.

In order to reduce systematic effects due to  $\gamma$'s
that are only partially contained and due to deviations from spherical symmetry, one should
select events near the detector's center. On the other hand,
statistics improves with increasing volume. We found that a good compromise
is to consider events such that the $\alpha$'s reconstructed positions are within
a sphere of 42 cm around the CTF center.
This sphere, after the cuts discussed above, contains $3.14 \cdot  10^4$ candidate decays.

The theoretical spectra have been produced with the CTF code,
described in \cite{Back:2004hd} and with the specific adjustments presented in the Appendix.

\subsection{\label{subsec:feedingProb}Feeding probability of the lowest state}
First we shall assume that the neutrino energy distribution $f(E_{\nu})$ is
given as a sum of ``universal'' functions, \ie,
\be
f(E_{\nu}) = \sum_n p_n F_{\mathrm{univ}} (E_{\nu}, Q-E_n) \quad ,
\ee
where $E_n$ is the energy of the $n$-th level ($E_0 = 0$), \ie,
the maximal energy which can be taken by the neutrino is $Q-E_n$,
and the functions $F_{\mathrm{univ}} (E_{\nu}, Q-E_n)$
are each normalized to unity.
The electron kinetic-energy distribution is:
\be
\phi(T_e) = \sum_n p_n \Phi_{\mathrm{univ}} (T_e , Q-E_n) \quad .
\ee
The universal distributions for neutrinos and electrons are related by:
\be
\Phi_{\mathrm{univ}} (T_e , Q-E_n) = F_{\mathrm{univ}} (E_{\nu} = Q-E_n - T_e, Q-E_n) \quad .
\ee
The populations of the 82 excited \nucleus[214]{Po} states are fixed at
the values given in the Table of Isotopes (ToI)~\cite{Akovali:1995},
apart for a common normalization factor, such that the total population
of these states is $(1- p_0)$.
This assumption for the excited states means that the relative intensities of the
$\gamma$ transition lines in the decay are exactly determined.

\begin{figure}
\includegraphics[width=0.6\columnwidth,angle=-90]{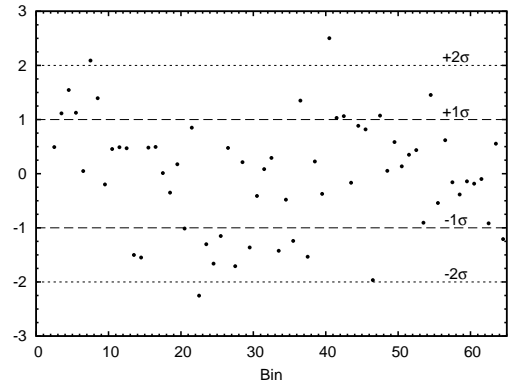}
 \caption[ggg]{Data and best fit: residuals. The values shown
 are the residual relative to figure~\ref{fig:dataBestFit}:
 data - best-fit value divided by the square root of data.
 The best fit is obtained with three parameters ($p_0$, light yield, and
 normalization) left as free for data in bins from 3 to 65.
\label{fig:residuals}
}
\end{figure}

We fitted the data with Montecarlo generated spectra leaving as free parameters:
\begin{enumerate}
  \item[(1)] $p_0$, the feeding probability of the lowest state;
  \item[(2)] the Light Yield $L$, defined as the number of photoelectrons that would be
  collected by 100 photomultipliers for an electron depositing 1~MeV at the center of CTF;
  \item[(3)] the normalization, \ie the number of reconstructed candidates
  (which should be equal to the number of candidates).
\end{enumerate}
The best fit function and the residuals are shown in
Figs.~\ref{fig:dataBestFit} and \ref{fig:residuals}.
At the minimum
$\chi^2 / \mathrm{d.o.f.} = 61.7 /(63-3)$, the  Light Yield
 $L = 321$~p.e./MeV~\footnote{Note that this Light Yield is within
 3\% the one determined at much lower energy by a fit to \nucleus[14]{C} events.},
 and the normalization factor is 0.998.
 The best fit value is $p_0 = 0.177$ with a statistical $1\sigma$  error of $\pm 0.004$.

%% figura 7

Systematic uncertainties arise from limitations of the Montecarlo simulation with
respect to the real detector (see Appendix).
We found that the largest uncertainties originate from the imperfect spherical
symmetry of the detector arising because of the deformations of the inner vessel (IV)
and to non-spherical distribution of the active PMTs.
The nature of these systematic errors makes them more important for events at
large radii or involving high-energy gammas, which can deposit energy
far away from the point of origin. Therefore, we estimated the effect of these
errors on our measurements by analyzing subgroup of events with different
distance $r$ from the center of the detector. Results show a consistent
behavior as function of $r$. Effects of the uncertainties on the quenching
parameters, subtraction of random coincidences, selection of the energy window,
and choice of the size of the energy binning have also been considered.
In conclusion the total systematic error is estimated as ${}^{+ 0.003}_{-0.001}$, so that:
\be
\mathrm{(CTF)}     \quad
p_0 = 0.177 \pm 0.004 (\mathrm{stat}) {}^{+ 0.003}_{-0.001} (\mathrm{sys}) \quad .
\ee
This value is consistent with that reported in
ToI~\cite{Akovali:1995}, $p_0(\mathrm{ToI}) = 0.182 \pm 0.006$.

\subsection{\label{subsec:shapeFactor}Shape factor for the pure beta transition}
Next we release the assumption that the spectrum for the transition to the
ground state has the ``universal shape''.
The electron energy distribution is assumed to be:
\be
\phi(T_e)= p_0 \Phi(T_e) + \sum_{n>0} p_n \Phi_{\mathrm{univ}} (T_e , Q-E_n)
\ee
where
\be
\label{eq:spectrumDistoY}
\Phi(T_e) = \Phi_{\mathrm{univ}} (T_e , Q)
\left( 1 + y \frac{T_e - \langle T_e \rangle}{\langle T_e \rangle} \right)
\ee
and the average energy $\langle T_e \rangle$
is  calculated over $\Phi_{\mathrm{univ}} (T_e , Q)$.

The dimensionless ``shape parameter'' $y$ describes thus a deviation
from the universal formula. Note that this simple parameterization
does not change the normalization of the distribution; it only changes its shape.
Other parameterizations are of course possible: at the level of the
accuracy of this preliminary study, they would not change our conclusions.

\begin{figure}
\includegraphics[width=0.6\columnwidth,angle=-90]{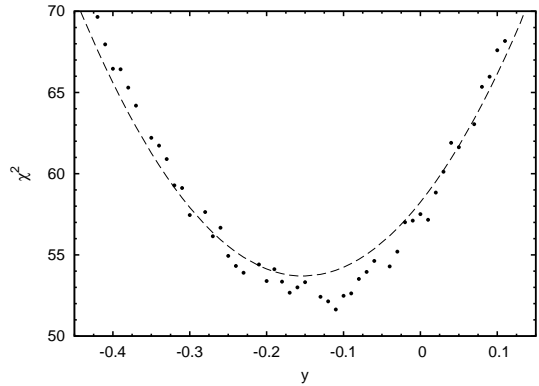}
 \caption[hhh]{The $\chi^2$ as a function of the deformation parameter $y$. The
 number of degrees of freedom is $(65-5)$.
\label{fig:deformationChi}
}
\end{figure}

Present data do not allow to independently determine
$p_0$, $p_1$, and $y$. We consider therefore as inputs the values given
in ToI, $p_0 = 0.182 \pm 0.006$ and $p_1 = 0.017 \pm 0.006$,
with errors assumed uncorrelated, and leave only $y$ as an unconstrained parameter.
The resulting $\chi^2$ is shown in
Fig.~\ref{fig:deformationChi} as a function of $y$.
At the best fit we find  $\chi^2 / \mathrm{d.o.f.} = 51.6 / (65-5)$,
$p_0 = 0.177$, $p_1 = 0.008$ and:
\be
\label{eq:Ybest}
y = -0.11  \pm 0.06 (\mathrm{stat})  \quad .
\ee

%% figura 8

The universal spectrum, $y = 0$, has a $\chi^2$ larger by 5.9 with respect
to the minimum: the statistical evidence for a deformed spectrum is about
$2.4\sigma$. If both $p_0$ and $p_1$ are left completely unconstrained,
one again finds that the best fit value for $y$ is $-0.11$, but with a
larger $1\sigma$  interval: $-0.53 < y < -0.09$. However, spectral deformation
is very sensitive to the lowest part of the visible energy and, therefore,
we expect larger systematic uncertainties than the one estimated in the case
of $p_0$. In fact, the analysis of subgroup of events with different
distances from the center of the detector, as we did for the feeding
probabilities, results in an estimated systematic error for $y$ of the same size
of its deviation from zero, \ie $\approx 0.10$.

Our present result only shows that the method is sensitive to the form of the
spectrum and has the potentiality of detecting spectral deformations.
However, interesting results can be obtained by achievable reductions
of statistical and systematic errors.

\subsection{\label{subsec:implicationSignal}Implications for the specific signal}
The geo-neutrino signal, $s(\nucleus[214]{Bi})$,
can be written as the sum of two contributions:
\be
\label{eq:geoSignalBi}
s(\nucleus[214]{Bi}) =  p_0 \langle\sigma\rangle_0 +  p_1 \langle\sigma\rangle_1 \quad,
\ee
where the cross section of reaction~(\ref{eq:inverseBeta}) is averaged over the
neutrino energy distribution. Assuming  universal shape, \ie,
 $\langle\sigma\rangle_0  = \int dE_{\nu}  \sigma(E_{\nu})
 F_{\mathrm{univ}} (E_{\nu}, Q-E_n) $,
 the cross sections are $\langle\sigma\rangle_0 = 7.76$ and $\langle\sigma\rangle_1 = 2.825$
  in units of $10^{-44}$~cm$^2$, with errors of order half of percent.
  Clearly, the largest contribution to the geo-neutrino signal is given by the
  first term in Eq.~(\ref{eq:geoSignalBi}), so that the relative error is practically
  the one on $p_0$.
  From the previous analysis we find:
\begin{eqnarray}
\label{eq:signalCTF}
\mathrm{(CTF)} &&\\ s(\nucleus[214]{Bi}) &=&
    \left( 1.42 \pm 0.03 (\mathrm{stat}) {}^{+ 0.023}_{-0.008} (\mathrm{sys})
    \right)\cdot 10^{-44}\mathrm{\ cm}^2 \quad .\nonumber
\end{eqnarray}
This should be compared with the result derived using the Table of Isotopes:
\begin{eqnarray}
\label{eq:signalToI}
\mathrm{(ToI)}  &&     \\
s(\nucleus[214]{Bi}) &=&
    \left( 1.46 \pm 0.05 (\mathrm{stat})
    \right)\cdot 10^{-44}\mathrm{\ cm}^2 \quad . \phantom{{}^{+ 0.023}_{-0.008} (\mathrm{sys})}
    \nonumber
 \end{eqnarray}
If spectral distortion is allowed in the form of
Eq.~(\ref{eq:spectrumDistoY}), then  $\langle\sigma\rangle_0$ becomes
\be
\langle\sigma\rangle_0 \to \langle\sigma\rangle_0 + y \langle \Delta\sigma\rangle_0
\quad ,
\ee
where  $\langle\sigma\rangle_0 = 7.76$ and  $ \langle \Delta\sigma\rangle_0 = - 4.52$.

If we substitute the value of $y$  in Eq.~(\ref{eq:Ybest})
and the corresponding values for $p_0 = 0.177$ and $p_1=0.008$,
 we find:
\be
   s(\nucleus[214]{Bi}) =
    \left( 1.48 \pm 0.01 (\mathrm{stat}) \pm 0.03 (\mathrm{sys})
    \right)\cdot 10^{-44}\mathrm{\ cm}^2 \quad .
\ee

Note that, if we leave completely unconstrained the shape,
$p_0$, $p_1$, and $y$, the effects  of changes of shape and of $p_0$ on the signal are
anti-correlated: if the spectrum is deformed so that there are more (less) low-energy
electrons, the corresponding best fit value for $p_0$ is lower (higher).
For instance using the present parameterization, we can let $y$ span
from -0.64 to 0.13, finding the corresponding \emph{best} fit values for
$p_0$ e $p_1$ with no constraint (we disregard the fact that these \emph{best} fit values have often
too large $\chi^2$): while the values of $p_0$ go from 0.13 to 0.20,
the signal changes only by about $\pm 2\%$.
In other words, the resulting signal is weakly dependent on the shape factor.
\section{\label{sec:conclusion}Concluding remarks}
So far, we have estimated the \nucleus[214]{Bi} geo-neutrino specific signal by
using CTF data resulting from a limited radon contamination. Our estimate has
a comparable error with that derived from ToI, see Eqs.~(\ref{eq:signalCTF})
and (\ref{eq:signalToI}). We remark, however, that our method has two advantages:
\begin{enumerate}
  \item[(i)] the pure beta transition can be detected in CTF and its probability
  can be measured directly  (whereas in the study of gamma lines alone its existence
  was inferred from the fact that the gamma counts did not match with the
  expected number of decays and the probability was evaluated from this mismatch);
  \item[(ii)] one can check the validity of the universal shape approximation
  for the most important decay mode.
\end{enumerate}
A dedicated experiment makes sense, with the aim of reducing the relative statistical error
$\Delta p_0 / p_0$ to the level of the relative error on the cross section,
$\Delta  \langle\sigma\rangle_0 / \langle\sigma\rangle_0 \approx 0.5\%$.
This requires a statistics larger by a factor of about 20, or some $6\cdot 10^5$
selected events. At the same time, one has to reduce systematic errors on the
correspondence between measured light and released energy: the largest improvement
should be obtained by concentrating the source near the center of the detector.

Our preliminary results are of encouragement towards a series of dedicated
measurements, with suitable sources dissolved in the liquid scintillator and
placed in a vial near the CTF center. As an example, a \nucleus[222]{Rn}
 source with an initial activity of 5~Bq would be tolerable for CTF and
it would produce some $2\cdot 10^6$ decays in 11 days (2 lifetimes).
Electrons are stopped inside the vial and the scintillation light is propagated
within CTF through the quartz, whereas gamma conversion occurs inside the CTF
inner vessel. The inward facing PMTs outside the shroud can thus detect light
originating from both electrons and gammas. Along these lines,
one can  get a better estimate of the specific signal of \nucleus[214]{Bi},
 provide measurements of the other  signals,
 $s(\nucleus[234]{Pa})$ and  $s(\nucleus[212]{Bi})$, relevant for geo-neutrino
 studies, and, more generally,  measure  the energy spectra of neutrinos
 from long lived  heavy nuclei.

%\section*{Acknowledgments}
\acknowledgments
We are grateful for enlightening discussions and valuable comments
to E.~Bellotti, G.~Bellini, B.~Ricci, and C.~Tomei.
We thank the Borexino collaboration for providing data.

This work was partially supported by MIUR (Ministero
dell'Istruzione, dell'Universit\`a e della Ricerca) under
MIUR-PRIN-2006 project ``Astroparticle physics''.

\appendix*
\section{Montecarlo simulation of CTF detector}
\label{sec:appenMCctf}
Because light propagation in a large-volume scintillator detector involves complex mechanisms,
the precise modeling of the detector response requires that many phenomena are taken into
account. Among the most relevant issues, it is worth mentioning the wavelength dependence
of the processes involved in light propagation, the reflection/refraction at the
scintillator/water interface, and the light reflection on the concentrators.
The need to follow each of some 12000 photons emitted per 1-MeV electron
event makes tracing MC code very slow.

A fast and reliable code has been developed for the CTF detector and it is
briefly described in this appendix; more details can be found in~\cite{Back:2004hd}.
 The code takes advantage of using average parameters (such as light yield, energy
 resolution, and spatial reconstruction precision) obtained analyzing the detector's data.
 Optimal sets of data for calibrating, tuning and testing the code are the
 $\beta$-decay spectrum from \nucleus[14]{C} and the easily
 identifiable $\alpha$'s from the radon-chain decays. The code has two parts:
 the electron-gamma shower simulation (EG code) and the simulation of the
 registered charge and position (REG code).

The EG code generates events at a random position with random initial direction
 (for $\gamma$'s) and follows the gamma-electron shower using the EGS-4
 code~\cite{Nelson:1985}. The low-energy electrons and alphas are not
 propagated in the program and are considered to be point-like sources,
 located at the initial coordinates. The mean registered charge corresponding to
 the electron kinetic energy $T_e$ is calculated with
 \be
 Q_e(r) = A\cdot T_e \cdot f(k_B,T_e) f_R(r) \quad,
 \ee
 where $f_R(r)$ is a radial factor, which takes into account the dependence
 of the registered charge on the distance from the detector's center,
 and $f(k_B,T_e)$ is the ionization quenching factor for electrons; the normalization
 of these two factors has been chosen such that
 $f_R(0) = f(k_B, T_e=1\mathrm{\ MeV}) = 1$. The method used to obtain $f_R(r)$
  consists in studying the response for mono-energetic alpha's as a function of
  their radial position and it is described in~\cite{Derbin:2003}.

  For the PC-based scintillator (PC+PPO 1.5 g/l), the quenching factor
   $k_B = (1.7 \pm 0.1)\cdot 10^3$~cm MeV$^{-1}$ was found to satisfy
   experimental data~\cite{Bellini:2008zza}. This value agrees with the fit
   to high-statistics $\beta$ spectrum of \nucleus[14]{C}.
   The $\alpha$ particles from \nucleus[214]{Po} decay, which tag the
   Bi-Po events considered in our study, have energy of 7.69~MeV and
   are quenched to an equivalent $\beta$-energy (produce the same
   amount of light in the scintillator of an electron) of $751 \pm 7$~keV.
   In the set of data selected for the present work quenching was higher
   (and the light yield lower) than in Ref.~\cite{Bellini:2008zza}
   due to the presence of oxygen
   in the scintillator (the radon originated from atmospheric air).
   In fact, the 7.69~MeV alphas are found at the lower equivalent $\beta$
   energy, $E = 643$~keV, for events selected around the detector's center.
   The ratio of two energies can be used to scale the quenching $k_B$ factor:
   the adopted value is $k_B = 0.0195$~cm MeV$^{-1}$.

   The $\gamma$'s are propagated using the EGS-4 code. As soon as the
   $i$-th electron of energy $T_{e_{i}}$ appears inside the scintillator,
   the corresponding fraction of total registered charge is calculated:
 \be
 \Delta Q_i = A\cdot T_{e_{i}} \cdot f(k_B, T_{e_{i}}) f_R(r_i) \quad.
 \ee
The total mean collected charge is defined, when the $\gamma$ is discarded
by the EG code, as the sum of the individual deposits:
 \be
 Q_{\gamma} = \sum_i \Delta Q_i  \quad.
 \ee
The weighted position is assigned to the final $\gamma$:
\be
\textbf{r}_w = \frac{\sum_i \Delta Q_i \cdot \textbf{r}_i}{\sum_i \Delta Q_i } \quad,
 \ee
 where $\Delta Q_i$ is the charge deposited
 by the $i$-th electron at the position $ \textbf{r}_i$.

 Once the position and deposited charge of the event have been generated by the
 EG code as described above, the second part of the code (REG) randomly
 generates the corresponding number of photoelectrons registered at each PMT;
 it takes into account the proper geometrical factor and assumes Poissonian
 distributions of photoelectrons number at each PMT.

 Finally, the energy-dependent radial reconstruction is simulated. The reconstruction
 precision is assumed to be defined by the number of PMTs fired and dependent only
 on the distance from the detector's center (spherical symmetry). These two
 assumptions have been confirmed by measurements using artificial radon sources
 inserted in the CTF detector~\cite{Alimonti:1998nt,McCarty:2006jp}.

The main source of systematic errors in our study is the departure from spherical
symmetry of the detector due to deformations of the inner vessel (IV) and non-uniform
distribution of active PMT on the
spherical surface surrounding the scintillator. The IV, a $500\mu$m thick nylon bag
containing four tons of low density
($\approx 0.88$~g/cm$^3$) scintillator, is immersed in water.
The buoyancy forces are compensated by supporting strings, but the deformations
are not measured precisely and are not accounted for in the MC modeling.
The maximum radial deviations from the ideal sphere can be as big as $5\div10$~cm,
though in average the radius of the sphere is $R=100$~cm.
Another source of systematics of the same nature (absent
in an ideal spherical detector) is the position dependence of the light-collection
efficiency function, $f_R(r)$, which is assumed to depend only on the distance
from the center and not on all three coordinates. The nature of these systematic
errors makes them more important for events at large radii or involving
high-energy gammas, which can deposit energy far away from the point of origin.

%\newpage

%

%
%\begin{widetext}
%\end{widetext}
\end{document}